\documentclass[pra,superscriptaddress,showpacs,twocolumn,10pt]{revtex4}
\usepackage{amssymb}
\usepackage{graphicx}

\def\>{\rangle}

\begin{document}

\newtheorem{corollary}{Corollary}
\newtheorem{definition}{Definition}
\newtheorem{example}{Example}
\newtheorem{lemma}{Lemma}
\newtheorem{proposition}{Proposition}
\newtheorem{theorem}{Theorem}
\newtheorem{fact}{Fact}
\newtheorem{property}{Property}

\title{Efficiency of Deterministic Entanglement Transformation}

\author{Runyao Duan}
\email{dry02@mails.tsinghua.edu.cn}
\author{Yuan Feng}
\email{feng-y@tsinghua.edu.cn}
\author{Zhengfeng Ji}
\email{jizhengfeng98@mails.tsinghua.edu.cn}
\author{Mingsheng Ying}
\email{yingmsh@tsinghua.edu.cn}

\affiliation{State Key Laboratory of Intelligent Technology and Systems,
Department of Computer Science and Technology,
Tsinghua University, Beijing, China, 100084}

\date{\today}

\begin{abstract}
We prove that sufficiently many copies of a bipartite entangled
pure state can always be transformed into some copies of another
one with certainty by local quantum operations and classical
communication. The efficiency of such a transformation is
characterized by deterministic entanglement exchange rate, and it
is proved to be always positive and bounded from top by the
infimum of the ratios of Renyi's  entropies of source state and
target state.  A careful analysis shows that the deterministic
entanglement exchange rate cannot be increased even in the
presence of catalysts. As an application, we show that there can
be two incomparable states with deterministic entanglement
exchange rate strictly exceeding 1.
\end{abstract}

\pacs{03.67.Mn, 03.65.Ud}

\maketitle

\section{Introduction}
As a valuable resource in quantum information processing, quantum
entanglement has been widely used in quantum cryptography
\cite{BB84}, quantum superdense coding  \cite{BS92}, and quantum
teleportation \cite{BBC+93}. Consequently, it remains the subject
of interest at present after years of investigations. Since
quantum entanglement often exists between different subsystems of
a composite system shared by spatially separated parties, a
natural constraint on the manipulation of entanglement is that the
separated parties are only allowed to perform quantum operations
on their own subsystems and to communicate to each other
classically (LOCC). Using this restricted set of transformations,
the parties are often required to optimally manipulate the
nonlocal resources contained in the initial entangled state.

 A
central problem about quantum entanglement is thus to find the
conditions of when a given entangled state can be transformed into
another one via LOCC. This problem can be solved in two different,
but complementary, contexts: the finite regime and the asymptotic
regime. In asymptotic regime, Bennett \textit{et al.} proposed in
Ref. \cite{BBPS96} a reversible protocol which shows that infinite
copies of a given bipartite entangled pure state $|\psi_1\rangle$
can always be transformed by LOCC into another given bipartite
entangled pure state $|\psi_2\rangle$ with ratio
$\frac{H(|\psi_1\rangle)}{H(|\psi_2\rangle)}$, where
$H(|\psi\rangle)$ is the entropy of entanglement of
$|\psi\rangle$. The first important step of entanglement
transformation in the finite regime was made by Nielsen in Ref.
\cite{NI99}, where he presented the condition of two bipartite
entangled pure states $|\psi_1\rangle$ and $|\psi_2\rangle$ with
the property that $|\psi_1\rangle$ can be locally converted into
$|\psi_2\rangle$ deterministically. More precisely,  Nielsen
proved that the transformation $|\psi_1\rangle\rightarrow
|\psi_2\rangle$ can be achieved with certainty by LOCC if and only
if the Schmidt coefficient vector of $|\psi_1\rangle$ is majorized
by that of $|\psi_2\rangle$. Nielsen's result has been extended in
several ways to the case where deterministic local transformation
cannot be achieved
\cite{Vidal99,JP99a,VJN00,JP99,SRS02,DF03a,DF03,FDY03}. These
efforts also lead to the surprising phenomenon of entanglement
catalysis \cite{JP99}. Unlike the asymptotic regime, it has been
shown that during the entanglement manipulation some nonlocal
properties of the system are irreversibly lost, and that
entanglement does not behave as an additive resource in the finite
regime.

 This paper  considers an interesting problem which in a sense can be
thought of as a combination of the finite regime and the
asymptotic regime. Suppose that two parties share $m$ copies of
entangled pure state $|\psi_1\rangle$, and want to
deterministically transform them into some copies of another state
$|\psi_2\rangle$ by LOCC. Let $f(m)$ be the maximal number of
copies of $|\psi_2\rangle$ they can obtain. Then the deterministic
entanglement exchange rate $D(|\psi_1\rangle,|\psi_2\rangle)$ may
be defined as the optimal ratio $\frac{f(m)}{m}$, where $m$ ranges
over all positive integers. The main aim of this paper is to
evaluate  $D(|\psi_1\rangle,|\psi_2\rangle)$. This problem has
some features of the asymptotic regime in the sense that the
number of copies of  the source state is sufficiently large. On
the other hand, it also shares some properties with
transformations in the finite regime since they need the
transformation to be implemented with certainty.

 In this paper we mainly consider the case of bipartite entangled
 pure states. First, we are able to prove that the deterministic
 entanglement exchange rate $D(|\psi_1\rangle,|\psi_2\rangle)$ is
 always positive, which means that sufficiently many copies of
 an entangled pure state $|\psi_1\rangle$ always can be
 transformed into some copies of another entangled pure one. Second,
  we define the entropy ratio $R(|\psi_1\rangle,|\psi_2\rangle)$
  to be the infimum of ratios of Renyi's entropies of
  $|\psi_1\rangle$ and $|\psi_2\rangle$. Then it is shown that
  $D(|\psi_1\rangle,|\psi_2\rangle)$ is bounded from top by
  $R(|\psi_1\rangle,|\psi_2\rangle)$. Furthermore, we examine a
  special case. If the target state $|\psi_2\rangle$ is maximally
  entangled, then the upper bound can be achieved. Indeed, an
  analytical formula for calculating
  $D(|\psi_1\rangle,|\psi_2\rangle)$ is given.

  A somewhat surprising thing comes up when we consider the
  influence of catalysis on the deterministic entanglement
  exchange rate. It is
  demonstrated that $D(|\psi_1\rangle,|\psi_2\rangle)$ cannot be
  enhanced  even allowing extra entangled states to serve  as
  catalysts. In other words, entanglement catalysis has no effect
  on the deterministic entanglement exchange rate. Nevertheless, we show
  that a catalyst state is also useful since sometimes
  it can help us to obtain better lower bounds of deterministic
  entanglement exchange rate easily.

   As applications, we also present some concrete examples. In
   particular, we show an interesting phenomenon: there
  exist two states with deterministic entanglement exchange rate
  strictly larger than $1$ though they are incomparable under
  LOCC. More explicitly,  although $|\psi_1\rangle$ cannot be
  transformed into $|\psi_2\rangle$ directly under LOCC, sometimes
  it is still possible to transform $m$ copies of $|\psi_1\rangle$ into $n$ copies of
  $|\psi_2\rangle$ with $n>m$. In some sense, this phenomenon
  confirms that it is reasonable to use the notion of deterministic entanglement exchange
  rate to characterize the efficiency of deterministic transformation under LOCC.

The rest of the paper is organized as follows. In Sec. II, we
formally introduce  the notion of deterministic entanglement
exchange rate $D(|\psi_1\rangle,|\psi_2\rangle)$, and prove that
this quantity is positive and bounded from top by
$R(|\psi_1\rangle,|\psi_2\rangle)$. A formula of
$D(|\psi_1\rangle,|\psi_2\rangle)$ when $|\psi_2\rangle$ is
maximally entangled is also presented in Sec. II. Next, in Sec.
III, the relation between entanglement catalysis and deterministic
entanglement exchange rate is examined carefully. As applications,
some concrete examples are given in Sec. IV. We draw a brief
conclusion together with some open problems for further study in
Sec. V.

\section{deterministic entanglement exchange rate and entropy ratio}

Let $|\psi_1\rangle=\sum_{i=1}^n
\sqrt{\alpha_i}|i\rangle|i\rangle$ be an entangled pure state with
ordered Schmidt coefficients $\alpha_1\geq \alpha_2\geq \ldots\geq
\alpha_n\geq 0$. We use the symbol $\lambda_{\psi_1}$ to denote
the ordered Schmidt coefficient vector of $|\psi_1\rangle$, i.e.,
$\lambda_{\psi_1}=(\alpha_1,\ldots, \alpha_n)$, which is just an
$n$-dimensional probability vector. Let
$|\psi_2\rangle=\sum_{i=1}^m \sqrt{\beta_i}|i\rangle|i\rangle$ be
a pure state with ordered Schmidt coefficient vector
$\lambda_{\psi_2}=(\beta_1,\ldots,\beta_m)$. We say that
$\lambda_{\psi_1}$ is majorzied by $\lambda_{\psi_2}$, denoted by
$\lambda_{\psi_1}\prec \lambda_{\psi_2}$, if the sum of $l$
largest components of $\lambda_{\psi_1}$ is not greater than that
of $\lambda_{\psi_2}$ for each $l=1,\ldots, {\rm min}(m,n)$. We
write $|\psi_1\rangle\rightarrow |\psi_2\rangle$ if
$|\psi_1\rangle$ can be deterministically transformed  into
$|\psi_2\rangle $  under LOCC.

Using the above notations,  Nielsen's theorem \cite{NI99} can be
stated  as follows: $|\psi_1\rangle\rightarrow |\psi_2\rangle$ if
and only if $\lambda_{\psi_1}\prec \lambda_{\psi_2}$.

By Nielsen's theorem, to determine whether $|\psi_1\rangle$ can be
transformed into $|\psi_2\rangle$ under LOCC, it suffices to
check  ${\rm min}\{m,n\}$ inequalities.  If
$|\psi_2\rangle$ is a maximally entangled state, then we need only
to check one inequality.

\begin{lemma}\label{lemma1}\upshape
Let $|\psi_1\rangle$ be an entangled state with the largest
Schmidt coefficient $\alpha_1$, and let
$|\Phi_k\rangle=\frac{1}{\sqrt{k}}\sum_{i=1}^k|i\rangle|i\rangle$
be a maximally entangled state. Then $|\psi_1\rangle\rightarrow
|\Phi_k\rangle$ if and only if $\alpha_1\leq \frac{1}{k}.$
\end{lemma}
\textit{Proof.} Immediately from the definition of majorization
and Nielsen's theorem. \hfill $\square$\\

Suppose now that  $|\psi_1\rangle$ and $|\psi_2\rangle$ are two
entangled states, we define $f(m)$ by the  maximal positive
integer $n$ such that $|\psi_1\rangle^{\otimes m}$ can be
transformed into $|\psi_2\rangle^{\otimes n}$ by LOCC, i.e.,
\begin{equation}\label{deffm}
                 f(m)={\rm max}\{n: |\psi_1\rangle^{\otimes m}\rightarrow |\psi_2\rangle^{\otimes n}\}.
\end{equation}
If the set on the right-hand side of Eq.(\ref{deffm}) is empty, we
simply set $f(m)$ to zero. Now  the {\it deterministic
entanglement exchange rate} from $|\psi_1\rangle$ to
$|\psi_2\rangle$, denoted by $D(|\psi_1\rangle,|\psi_2\rangle)$,
is defined as the supremum of ratios of $f(m)$ and $m$ for any
positive integer $m$, i.e.,
\begin{equation}\label{defdr}
                 D(|\psi_1\rangle,|\psi_2\rangle)= {\rm sup}_{m\geq
                 1}\frac{f(m)}{m}.
\end{equation}
Intuitively, for a sufficiently large $m$, we can transform $m$
copies of $|\psi_1\rangle$ exactly into
$mD(|\psi_1\rangle,|\psi_2\rangle)$ copies of $|\psi_2\rangle$ by
LOCC.

It may be instructive to clarify  the difference between the
deterministic entanglement exchange rate introduced above and the
asymptotic entanglement exchange rate considered by Bennett $et\
al$.  Recall from \cite{BBPS96} that the asymptotic entanglement
exchange rate, denoted by
$E^{asy}(|\psi_1\rangle,|\psi_2\rangle)$, is given by
\begin{equation}\label{asymptotic}
E^{asy}(|\psi_1\rangle,|\psi_2\rangle)=\frac{H(|\psi_1\rangle)}{H(|\psi_2\rangle)},
\end{equation}
where $H(|\psi_1\rangle)=-\sum\alpha_i\log_2 \alpha_i$ is the
entropy of entanglement of $|\psi_1\rangle$. The quantity
$E^{asy}(|\psi_1\rangle,|\psi_2\rangle)$ has a nice physical
meaning. In fact, for a sufficiently large $m$, we can
approximately transform $m$ copies of $|\psi_1\rangle$ into
$mE^{asy}(|\psi_1\rangle,|\psi_2\rangle)$ copies of
$|\psi_2\rangle$. Here `approximately' means that the resulted
state is a good approximation of $|\psi_2\rangle^{\otimes
mE^{asy}(|\psi_1\rangle,|\psi_2\rangle)}$ in the sense that the
fidelity between them tends to one with high probability when $m$
tends to infinity. On the other hand, as we just mentioned, we can
transform $m$ copies of $|\psi_1\rangle$ exactly into
$mD(|\psi_1\rangle,|\psi_2\rangle)$ copies of $|\psi_2\rangle$ by
LOCC for a sufficiently large $m$. Thus, although both
$E^{asy}(|\psi_1\rangle,|\psi_2\rangle)$ and $D(|\psi_1\rangle,
|\psi_2\rangle)$ are defined in an asymptotic sense,  the former
characterizes  the efficiency of approximate entanglement
transformation, while the later represents the efficiency of
deterministic entanglement transformation. We will see later that
$E^{asy}(|\psi_1\rangle,|\psi_2\rangle)$ is an upper bound of
$D(|\psi_1\rangle, |\psi_2\rangle)$ and in some special cases
these two quantities  coincide.  In  the rest of present paper, we
investigate properties of the deterministic entanglement exchange
rate carefully.

 First, one can easily check that the deterministic entanglement exchange rate defined above has
the following interesting properties:

\begin{lemma}\label{dcrproperty}\upshape
If $|\psi_1\rangle$, $|\psi_2\rangle$ and $|\psi_3\rangle$ are
three entangled pure states, then

1. $D(|\psi_1\rangle^{\otimes p}, |\psi_2\rangle^{\otimes
q})=\frac{p}{q} D(|\psi_1\rangle, |\psi_2\rangle).$

2. $D(|\psi_1\rangle, |\psi_2\rangle)\cdot D(|\psi_2\rangle,
|\psi_3\rangle)\leq D(|\psi_1\rangle, |\psi_3\rangle)$.
Especially, $D(|\psi_1\rangle, |\psi_2\rangle)D(|\psi_2\rangle,
|\psi_1\rangle)\leq 1$.

3. $D(|\psi_1\rangle\otimes |\psi_2\rangle, |\psi_3\rangle)\geq
D(|\psi_1\rangle, |\psi_3\rangle)+D(|\psi_2\rangle,
|\psi_3\rangle)$.

4. $D(|\psi_1\rangle,|\psi_2\rangle)\geq 1$ and
$D(|\psi_2\rangle,|\psi_1\rangle)\geq 1$ if and only if
$\lambda_{\psi_1}=\lambda_{\psi_2}$.

\end{lemma}

It would be desirable to know the precise value of
$D(|\psi_1\rangle,|\psi_2\rangle)$. However,  unlike the
asymptotic entanglement exchange rate,  we still don't know how to
compute the deterministic entanglement exchange rate at present.
Nevertheless, we can obtain a lower bound and an upper bound of
$D(|\psi_1\rangle,|\psi_2\rangle)$, respectively.

Before proving these two bounds, let us review some elements
of Renyi's entropy \cite{Ren84}. Recall that the $\tau$-order
Renyi's entropy of $|\psi_1\rangle$ is defined by
\begin{equation}\label{renyi}
S^{(\tau)}(|\psi_1\rangle)=\frac{1}{1-\tau}\log_2
\sum_{i=1}^n\alpha_i^{\tau}, {\rm \ for\  any\ }\tau >0 {\rm\ and\
} \tau\neq 1,
\end{equation}
where $(\alpha_1,\cdots,\alpha_n)$ is the ordered Schmidt
coefficient vector of $|\psi_1\rangle$. For the sake of
convenience, let $S^{(0)}(|\psi_1\rangle)=\log_2 d$,
$S^{(1)}(|\psi_1\rangle)=H(|\psi_1\rangle)$, and
$S^{(+\infty)}(|\psi_1\rangle)=-\log_2\alpha_1$, where $d$ is the
number of non-zero Schmidt coefficients of $|\psi_1\rangle$. It is
easy to verify that $S^{(\tau)}(|\psi\rangle)$ is continuous and
bounded  for any  $\tau\in [0,+\infty]$.

Renyi's entropy enjoys many useful properties.
The most interesting one is the additivity under tensor product.
That is,  $S^{(\tau)}(|\psi_1\rangle\otimes
|\psi_2\rangle)=S^{(\tau)}(|\psi_1\rangle)+S^{(\tau)}(|\psi_2\rangle)$.
Especially, $S^{(\tau)}(|\psi_1\rangle^{\otimes
m})=mS^{(\tau)}(|\psi_2\rangle)$ for any positive integer $m$. It
is also worth noting that Renyi's entropy does not increase under LOCC.
So, $|\psi_1\rangle\rightarrow |\psi_2\rangle$ implies
$S^{(\tau)}(|\psi_1\rangle)\geq S^{(\tau)}(|\psi_2\rangle)$.

Now we can use Renyi's entropy to define a quantity
$R(|\psi_1\rangle,|\psi_2\rangle)$ as follows:
\begin{equation}\label{renyiratio}
R(|\psi_1\rangle,|\psi_2\rangle)={\rm inf}_{\tau \geq
0}\frac{S^{(\tau)}(|\psi_1\rangle)}{S^{(\tau)}(|\psi_2\rangle)}.
\end{equation}
That is, $R(|\psi_1\rangle,|\psi_2\rangle)$ is the infimum of the
ratios of the Renyi's entropies of $|\psi_1\rangle$ and
$|\psi_2\rangle$. We name this useful quantity the `entropy ratio' of $|\psi_1\rangle$ and $|\psi_2\rangle$.

It is easy to prove  that $R(|\psi_1\rangle,|\psi_2\rangle)$ has
the following properties:

\begin{lemma}\label{renyiproperty}\upshape
If $|\psi_1\rangle$, $|\psi_2\rangle$ and $|\psi_3\rangle$ are
three entangled pure states, then

 1. $R(|\psi_1\rangle^{\otimes p}, |\psi_2\rangle^{\otimes
q})=\frac{p}{q}R(|\psi_1\rangle, |\psi_2\rangle).$

2. $R(|\psi_1\rangle, |\psi_2\rangle)R(|\psi_2\rangle,
|\psi_3\rangle)\leq R(|\psi_1\rangle, |\psi_3\rangle)$.
Especially, $R(|\psi_1\rangle,|\psi_2\rangle)
R(|\psi_2\rangle,|\psi_1\rangle)\leq 1$.

3. $R(|\psi_1\rangle\otimes |\psi_2\rangle, |\psi_3\rangle)\geq
R(|\psi_1\rangle, |\psi_3\rangle)+R(|\psi_2\rangle,
|\psi_3\rangle)$.

4. $R(|\psi_1\rangle,|\psi_2\rangle)\geq 1$ and
$R(|\psi_2\rangle,|\psi_1\rangle)\geq 1$ if and only if
$\lambda_{\psi_1}=\lambda_{\psi_2}$.

5. there exists $\tau_0\in [0,+\infty]$ such that
$R(|\psi_1\rangle,|\psi_2\rangle)=\frac{S^{(\tau_0)}(|\psi_1\rangle)}{S^{(\tau_0)}(|\psi_2\rangle)}$.
\end{lemma}

Comparing Lemmas \ref{dcrproperty} and \ref{renyiproperty}, one can see
that $R(|\psi_1\rangle,|\psi_2\rangle)$ and
 $D(|\psi_1\rangle,|\psi_2\rangle)$ enjoy many similar
properties. Indeed, the former serves as an upper bound on the
latter, as the following theorem indicates:
\begin{theorem}\label{measure}\upshape
If $|\psi_1\rangle$ and $|\psi_2\rangle$ are two entangled states,
then
$$0<D(|\psi_1\rangle,|\psi_2\rangle)\leq
R(|\psi_1\rangle,|\psi_2\rangle).$$
\end{theorem}

\textit{Proof.} To prove the first inequality, we only need to
show that for some positive integer $m$, $|\psi_1\rangle^{\otimes
m}\rightarrow |\psi_2\rangle$, which yields
$D(|\psi_1\rangle,|\psi_2\rangle)\geq \frac{1}{m}>0$. Without loss
of generality, let $|\psi_2\rangle$  be  an $n\times n$ state, and
$|\Phi_n\rangle$ an $n\times n$ maximally entangled state. It is
obvious that $|\Phi_n\rangle\rightarrow |\psi_2\rangle$. We shall
show that by a careful choice of $m$ we have
$|\psi_1\rangle^{\otimes m}\rightarrow |\Phi_n\rangle$, thus
$|\psi_1\rangle^{\otimes m}\rightarrow |\psi_2\rangle$. In fact,
let $\alpha_1$ be the maximal component of $\lambda_{\psi_1}$.
Then, since $|\psi_1\rangle$ is an entangled state, it follows
that $0<\alpha_1<1$. Hence it is always possible to take an $m$
such that $\alpha_1^m\leq \frac{1}{n}$. Applying Lemma
\ref{lemma1} leads us to $|\psi_1\rangle^{\otimes m}\rightarrow
|\Phi_n\rangle$.

To deal with the second inequality, we utilize the above argument which states that
for some sufficiently large $m$,
$|\psi_1\rangle^{\otimes m}\rightarrow |\psi_2\rangle^{\otimes
f(m)}$ and $f(m)\geq 1$. By the properties of Renyi's entropy
mentioned above, it follows that $ mS^{(\tau)}(|\psi_1\rangle)\geq
f(m)S^{(\tau)}(|\psi_2\rangle)$, or
\begin{equation}\label{eq2}
           \frac{f(m)}{m}\leq
           \frac{S^{(\tau)}(|\psi_1\rangle)}{S^{(\tau)}(|\psi_2\rangle)}.
\end{equation}
Then the second inequality holds by taking supremum according to
$m$ on the left-hand side and infimum according to $\tau$ on the
right-hand side of the above formula, respectively.
\hfill $\square$\\

By the definition of entropy ratio, we have
$R(|\psi_1\rangle,|\psi_2\rangle)\leq E^{asy}(|\psi_1\rangle,
|\psi_2\rangle)$. Hence an immediate consequence of Theorem
\ref{measure} is that $D(|\psi_1\rangle,|\psi_2\rangle)\leq
E^{asy}(|\psi_1\rangle,|\psi_2\rangle)$. In other words, the
asymptotic entanglement exchange rate considered in Ref.
\cite{BBPS96} serves as an upper bound of the deterministic
entanglement exchange rate.

Theorem \ref{measure} deserves some more remarks. First,
$D(|\psi_1\rangle,|\psi_2\rangle)>0$ reveals a fundamental
property of entangled pure states. That is,  any two entangled
pure states are interconvertible in the sense that  sufficiently
many copies of one state can always be exactly transformed  into
some copies of another state by LOCC \cite{JDY04}. Although this
seems very reasonable, it is not all obvious that it should be the
case. Since it is well known that for mixed states, there exist
bounded entangled states that cannot be concentrated into a
singlet even asymptotically \cite{HHH98}. Moreover, as shown in
Ref. \cite{KENT98}, the maximal conversion probability of a
generic mixed state to an entangled pure state is always  zero.
Thus entangled pure states can be treated as the most valuable
entanglement resources, and they are interconvertible under LOCC.
Second, the theorem also indicates that $R(|\psi_1\rangle,
|\psi_2\rangle)$ is an upper bound of
$D(|\psi_1\rangle,|\psi_2\rangle)$. Whether this bound is tight or
not is still unknown. In the following, we shall further
investigate the property of $D(|\psi_1\rangle,|\psi_2\rangle)$,
and it will be shown that in some special but interesting cases,
this upper bound can be achieved.

In particular, if target state is maximally entangled, we are able
to calculate deterministic entanglement exchange rate explicitly,
which coincides with the upper bound presented above.
\begin{theorem}\label{ratebell}\upshape
If $|\psi_1\rangle$ is an entangled  state with the greatest
Schmidt coefficient $\alpha_1$,  and $|\Phi_k\rangle$ is a
$k\times k$ maximally entangled state, then $D(|\psi_1\rangle,
|\Phi_k\rangle)=-\log_k \alpha_1$.
\end{theorem}

\textit{Proof.} By Theorem \ref{measure},  for a sufficiently
large positive integer $m$, it holds that $f(m)\geq 1$. Moreover,
by the definition of $f(m)$, it follows that
\begin{equation}\label{maj1}
 |\psi_1\rangle^{\otimes m}\rightarrow |\Phi_k\rangle^{\otimes f(m)}
\end{equation}
but \begin{equation}\label{maj2} |\psi_1\rangle^{\otimes
m}\nrightarrow |\Phi_k\rangle^{\otimes f(m)+1}.
\end{equation}
From Lemma \ref{lemma1}, Eqs. (\ref{maj1}) and (\ref{maj2}) are
equivalent to
\begin{equation}\label{bellmaj1}
             \left (\frac{1}{k}\right )^{f(m)+1}<\alpha_1^m\leq
             \left (\frac{1}{k}\right)^{f(m)},
\end{equation}
or
\begin{equation}\label{bellmaj2}
             -\log_k\alpha_1-\frac{1}{m}<\frac{f(m)}{m}\leq
             -\log_k\alpha_1.
\end{equation}
With $m$ tending to $+\infty$, we have $D(|\psi_1\rangle,
|\Phi_k\rangle)=-\log_k \alpha_1$. \hfill $\square$\\

We notice that the problem of deterministic concentration of Bell
pairs from a finite number of partially entangled pairs was
considered in Ref. \cite{FM01}. As a natural extension of the
solution of problem, the quantity $-\log_k \alpha_1$ was treated
there as an entanglement measure of state $|\psi_1\rangle$. It is
clear that the precise meaning of this quantity is the
deterministic entanglement exchange rate
$D(|\psi_1\rangle,|\Phi_k\rangle)$.

Except for some trivial cases, the transformations in the finite
regime is always irreversible in the sense some entanglement is
lost during the manipulation \cite{NI99, Vidal99, FM01, SRS02}.
Interestingly, if  sufficiently many copies of source state are
available, sometimes we may do entanglement transformation
deterministically without loss of entanglement. For example, by
Theorem \ref{ratebell},
\begin{equation}\label{belltobell}
D(|\Phi_{k_1}\rangle,|\Phi_{k_2}\rangle)=\frac{1}{D(|\Phi_{k_2}\rangle,|\Phi_{k_1}\rangle)}=
\log_{k_2}k_1.
\end{equation}
In other words, if  both  source state and target state are
maximally entangled, then the deterministic entanglement exchange
rate coincides with the asymptotic entanglement entanglement
exchange rate. Thus in this special case the transformation can be
reversible.

\section{entanglement catalysis and deterministic entanglement exchange rate}
 In this section, we examine the relation between
 entanglement catalysis and deterministic entanglement exchange rate.
 More precisely, we will answer the following question: can the
 deterministic entanglement exchange rate be increased by introducing catalysts?

We say that $|\psi_1\rangle$ can be catalyzed into
$|\psi_2\rangle$ if there exists a state $|\phi\rangle$ such that
$|\psi_1\rangle\otimes |\phi\rangle\rightarrow
|\psi_2\rangle\otimes |\phi\rangle$. This kind of transformation
is often called {\it entanglement-assisted local transformation},
abbreviated by ELOCC \cite{JP99}. And the state $|\phi\rangle$ is
called a catalyst for the transformation.   Since an ELOCC
transformation  is always not less, and sometimes
strictly more, powerful than a LOCC transformation, the
deterministic entanglement exchange rate may be increased by allowing
extra states to serve as catalysts. However, we shall prove that
it is not the case.

To be concise, we define the notion of
   deterministic entanglement exchange rate under ELOCC. More precisely,
suppose that  $|\psi_1\rangle$ and $|\psi_2\rangle$ are two given
states, we define $f'(m)$ as the  maximum $n$ such that
$|\psi_1\rangle^{\otimes m}$ can be catalyzed into
$|\psi_2\rangle^{\otimes n}$. That is,
\begin{equation}\label{deffm1}
                 f'(m)={\rm max}\{n: \exists |\phi\rangle {\rm\ s.t.\ }|\psi_1\rangle^{\otimes m}\otimes |\phi\rangle\rightarrow |\psi_2\rangle^{\otimes n}\otimes |\phi\rangle\}.
\end{equation}
If the set on the right-hand side of Eq.(\ref{deffm1}) is empty,
we simply set $f'(m)$ to zero. Now  the {\it
entanglement-assisted deterministic entanglement exchange rate} from
$|\psi_1\rangle$ to $|\psi_2\rangle$ can be defined as
\begin{equation}\label{defdr1}
                 D^c(|\psi_1\rangle,|\psi_2\rangle)= {\rm sup}_{m\geq
                 1}\frac{f'(m)}{m},
\end{equation}
where the superscript $c$ denotes ``catalyst-assisted
transformation". Intuitively, for a sufficiently large $m$,
$mD^c(|\psi_1\rangle,|\psi_2\rangle)$ denotes the maximal number
of the copies $|\psi_2\rangle$ that can be deterministically
obtained  from $m$ copies of $|\psi_1\rangle$ by ELOCC.

Now the relation between entanglement catalysis and deterministic
entanglement exchange rate is summarized by the following:

\begin{theorem}\label{catalystrate}\upshape
   If  $|\psi_1\rangle$ and $|\psi_2\rangle$ are two entangled
   states, then
   $D^c(|\psi_1\rangle,|\psi_2\rangle)=D(|\psi_1\rangle,|\psi_2\rangle)$.
\end{theorem}

In other words, deterministic entanglement exchange rate cannot be
increased by entanglement-assisted transformation. The proof we
will give in the following indicates that this result also holds
in the multipartite setting though the existence of multipartite
catalyst is still unknown.

 To prove the
above theorem, we need a useful lemma. This lemma also shows some
connection between entanglement catalysis and the deterministic
entanglement exchange rate.

\begin{lemma}\label{sufficent}\upshape
  If $|\psi_1\rangle\rightarrow|\psi_2\rangle$ under ELOCC, then $D(|\psi_1\rangle,|\psi_2\rangle)\geq 1$.
\end{lemma}
 \textit{Proof.} By assumption, there exists a state
$|\phi\rangle$ such that
\begin{equation}\label{elocc}
|\psi_1\rangle\otimes |\phi\rangle\rightarrow
|\psi_2\rangle\otimes
  |\phi\rangle.
\end{equation}
From Theorem \ref{measure}, we can find a constant $m_0$ such that
\begin{equation}\label{gcatalyst}
|\psi_1\rangle^{\otimes m_0 }\rightarrow
  |\phi\rangle.
\end{equation}

Now suppose that we have $m$ copies of $|\psi_1\rangle$, where
$m>m_0$. The following protocol shows that $f(m)\geq m-m_0$:

Step 1. perform  $|\psi_1\rangle^{\otimes m}\rightarrow
|\psi_1\rangle^{\otimes (m-m_0)}\otimes |\phi\rangle$;

Step 2. perform $|\psi_1\rangle^{\otimes (m-m_0)}\otimes
|\phi\rangle\rightarrow |\psi_2\rangle^{\otimes (m-m_0)}\otimes
|\phi\rangle$.

Step 1 is a simple use of Eq. (\ref{gcatalyst}), and step 2 can be
realized by using Eq. (\ref{elocc}) $(m-m_0)$ times.

By the definition of $D(|\psi_1\rangle,|\psi_2\rangle)$, it
follows that
$$D(|\psi_1\rangle,|\psi_2\rangle)={\rm sup}_{m\geq 1}\frac{f(m)}{m}\geq {\rm sup}_{m>m_0}\frac{m-m_0}{m}=1.$$
That completes the proof of this lemma. \hfill $\square$\\

{\it Proof of Theorem \ref{catalystrate}.} For a sufficiently
large $m$, we have  $|\psi_1\rangle^{\otimes m}\rightarrow
|\psi_2\rangle^{\otimes f'(m)}$ under ELOCC. Thus, by Lemma
\ref{sufficent}, it follows that
$$D(|\psi_1\rangle^{\otimes m}, |\psi_2\rangle^{\otimes
f'(m)})\geq 1.$$ Furthermore, from (1) of Lemma \ref{dcrproperty},
we have
$$D(|\psi_1\rangle,|\psi_2\rangle)=\frac{f'(m)}{m}D(|\psi_1\rangle^{\otimes m}, |\psi_2\rangle^{\otimes
f'(m)}).$$ Combining the above two equations, we derive
$$D(|\psi_1\rangle,|\psi_2\rangle)\geq \frac{f'(m)}{m}$$
for any positive integer $m$. Taking supremum according to $m$
yields
$$D(|\psi_1\rangle,|\psi_2\rangle)\geq
D^c(|\psi_1\rangle,|\psi_2\rangle).$$ On the other hand, it is
obvious that $$D(|\psi_1\rangle,|\psi_2\rangle)\leq
D^c(|\psi_1\rangle,|\psi_2\rangle).$$ That completes our proof.
\hfill $\square$\\

As a direct implication of Theorems \ref{measure} and
\ref{catalystrate}, we have the following:
\begin{corollary}\label{edis1}\upshape
 If  $|\psi_1\rangle\rightarrow |\psi_2\rangle$ under ELOCC and $R(|\psi_1\rangle, |\psi_2\rangle)= 1$, then
$D(|\psi_1\rangle,|\psi_2\rangle)=1$.
\end{corollary}

Some special cases of Corollary \ref{edis1} are of great interest.
Suppose that $|\psi_1\rangle\rightarrow
|\psi_2\rangle$ by ELOCC, then $D(|\psi_1\rangle,|\psi_2\rangle)=1$ if these two states share the same
greatest Schmidt coefficient or if they have the same number of
nonzero Schmidt coefficients.

\section{Some Applications}
In this section, we give some concrete examples. First, the
problem of calculating deterministic entanglement exchange rate of
two $2\times 2$ entangled pure states is considered. We present
some partial results about this problem in the following example.
\begin{example}\label{example1}\upshape Let
$|\psi_1\rangle=\sqrt{p}|00\rangle+\sqrt{1-p}|11\rangle$ and
$|\psi_2\rangle=\sqrt{q}|00\rangle+\sqrt{1-q}|11\rangle$, where
$\frac{1}{2}<p,q<1$. Our aim here is to calculate
$D(|\psi_1\rangle,|\psi_2\rangle)$.

    If $p\leq q$, then it is easy to verify that
$|\psi_1\rangle\rightarrow |\psi_2\rangle$ and $R(|\psi_1\rangle,
|\psi_2\rangle)=1$. Thus $D(|\psi_1\rangle, |\psi_2\rangle)=1$.

The case of $p>q$ is much more complicated and it  seems too
difficult to give  a precise expression of $D(|\psi_1\rangle,
|\psi_2\rangle)$. We consider a special case here. If there is
some positive real $\mu$ such that $p=\mu^{m}$ and $q=\mu^{n}$,
where $m=1,2,3$ and $n>m$. Then
$D(|\psi_1\rangle,|\psi_2\rangle)=\frac{m}{n}$.

In fact, under the assumptions, a direct calculation carries out
that $|\psi_1\rangle^{\otimes n}\rightarrow
|\psi_2\rangle^{\otimes m}$, thus
$D(|\psi_1\rangle,|\psi_2\rangle)\geq \frac{m}{n}$. On the other
hand,
$$R(|\psi_1\rangle,|\psi_2\rangle)\leq
\frac{S^{(+\infty)}(|\psi_1\rangle)}{S^{(+\infty)}(|\psi_2\rangle)}=\frac{m}{n}.$$
By Theorem \ref{measure}, it follows that
$D(|\psi_1\rangle,|\psi_2\rangle)=\frac{m}{n}$.
 \hfill
$\square$
\end{example}

  Nielsen's theorem implies that there are incomparable states
  $|\psi_1\rangle$ and $|\psi_2\rangle$ in the sense neither $|\psi_1\rangle\rightarrow
  |\psi_2\rangle$ nor $|\psi_2\rangle\rightarrow |\psi_1\rangle$.
  Thus the maximal conversion probability between two incomparable
  state is strictly less than $1$.  The well-known effects of entanglement
  catalysis or multiple-copy entanglement transformation \cite{SRS02, DF03a,DF03} can help
  us to transform some copies of source state $|\psi_1\rangle$ into
  the same number of copies of target state $|\psi_2\rangle$. We
  further ask: can we obtain more copies of target state than
  source state?  To one's surprise, the answer for this question is yes.
  Specifically, the following example indicates that there can be incomparable
  states $|\psi_1\rangle$ and $|\psi_2\rangle$ such that $D(|\psi_1\rangle,
  |\psi_2\rangle)>1$. It also shows that sometimes  a catalyst state can
help us to obtain a more precise lower bound of  deterministic
entanglement exchange rate.
\begin{example}\label{example2}\upshape
 Let $|\psi_1\rangle=\sqrt{0.40}|00\rangle+\sqrt{0.36}|11\rangle+\sqrt{0.14}|22\rangle+\sqrt{0.10}|33\rangle
$  and
$|\psi_2\rangle=\sqrt{0.50}|00\rangle+\sqrt{0.25}|11\rangle+\sqrt{0.25}|22\rangle
$.
 It is obvious that $|\psi_1\rangle\nrightarrow |\psi_2\rangle$.
However,  one can easily check that $|\psi_1\rangle^{\otimes
k}\rightarrow |\psi_2\rangle^{\otimes k}$ for any $k\geq 2$. This
is just the effect of multiple-copy entanglement transformation.
The most interesting thing here is that $|\psi_1\rangle^{\otimes
8}\rightarrow |\psi_2\rangle^{\otimes 9}$ by Nielsen's theorem.
Thus, $D(|\psi_1\rangle,|\psi_2\rangle)\geq 9/8$.

 By Theorem
\ref{catalystrate}, we have known that entanglement catalysis
cannot increase the deterministic entanglement exchange rate. But
a catalyst state is still useful in the sense that it can help us
to obtain more precise lower bound of
$D(|\psi_1\rangle,|\psi_2\rangle)$. In fact,  since
$|\psi_1\rangle^{\otimes 7}$ and $|\psi_2\rangle^{\otimes 8}$ are
LOCC incomparable, we seek for a potential catalyst to help the
transformation. Taking
$|\phi\rangle=\sqrt{0.60}|44\rangle+\sqrt{0.40}|55\rangle$, by a
routine calculation, we have $$|\psi_1\rangle^{\otimes 7}\otimes
|\phi\rangle^{\otimes 4}\rightarrow |\psi_2\rangle^{\otimes
8}\otimes|\phi\rangle^{\otimes 4},$$ where $|\phi\rangle$ is
called a multiple-copy catalyst for the transformation from
$|\psi_1\rangle$ to $|\psi_2\rangle$  \cite{DF03a}. Thus
$D(|\psi_1\rangle,|\psi_2\rangle)\geq 8/7$  from Theorem
\ref{catalystrate}. So, a  more precise lower bound of
$D(|\psi_1\rangle, |\psi_2\rangle)$ is obtained.

From the above argument, one may naturally ask: is it possible to
obtain a more precise lower bound of
$D(|\psi_1\rangle,|\psi_2\rangle)$ by transforming $6$ copies of
$|\psi_1\rangle$ into 7 copies of $|\psi_2\rangle$? Unfortunately,
since $$R(|\psi_1\rangle,|\psi_2\rangle)\leq
\frac{S^{(2)}(|\psi_1\rangle)}{S^{(2)}(|\psi_2\rangle)}=1.1643<7/6,$$
it is impossible to transform $6$ copies of $|\psi_1\rangle$ into
$7$ copies of $|\psi_2\rangle$ by Theorem \ref{measure}. \hfill
$\square$
\end{example}

A catalyst state may also help us to achieve the upper bound of
$D(|\psi_1\rangle,|\psi_2\rangle)$ easily. We demonstrate this
point in the following example.
\begin{example}\label{example3}\upshape
Take  source state and target state as
$|\psi_1\rangle=\frac{1}{\sqrt{1.01}}(\sqrt{0.4}|00\rangle+\sqrt{0.4}|11\rangle+\sqrt{0.1}|22\rangle+\sqrt{0.1}|33\rangle+\sqrt{0.01}|44\rangle)$
and
$|\psi_2\rangle=\frac{1}{\sqrt{1.01}}(\sqrt{0.5}|00\rangle+\sqrt{0.25}|11\rangle+\sqrt{0.2}|22\rangle+\sqrt{0.05}|33\rangle+\sqrt{0.01}|44\rangle)$,
respectively.

By Remark 1 in Ref. \cite{FDY03}, we have
$|\psi_1\rangle^{\otimes m}\nrightarrow |\psi_2\rangle^{\otimes
m}$ for any positive integer $m$. Thus $f(m)<m$ for any $m$, which
yields $D(|\psi_1\rangle, |\psi_2\rangle)\leq 1$ and any finite
$m$ cannot attain the upper bound $R(|\psi_1\rangle,
|\psi_2\rangle)=1$. Now if we take
$|\phi\rangle=\sqrt{0.6}|55\rangle+\sqrt{0.4}|66\rangle$, then a
simple calculation carries out that $|\phi\rangle$ is a
multiple-copy catalyst for the transformation from
$|\psi_1\rangle$ to $|\psi_2\rangle$, since it holds that
$$|\psi_1\rangle\otimes |\phi\rangle^{\otimes 11}\rightarrow
|\psi_2\rangle\otimes |\phi\rangle^{\otimes 11}.$$ Applying Lemma
\ref{sufficent}, we have $D(|\psi_1\rangle,|\psi_2\rangle)\geq 1$.
Therefore, it holds that $D(|\psi_1\rangle,|\psi_2\rangle)=1$, and
this value can be attained by transforming $|\psi_1\rangle$ into
$|\psi_2\rangle$ with the aid of the catalyst state
$|\phi\rangle^{\otimes 11}$.\hfill $\square$
\end{example}

The last example is aimed to demonstrate the difference between
probabilistic transformation and deterministic transformation.

\begin{example}\label{example4}\upshape
  Take
  $|\psi_1\rangle=\sqrt{0.4}|00\rangle+\sqrt{0.4}|11\rangle+\sqrt{0.2}|22\rangle$
  and $|\psi_2\rangle=\sqrt{0.5}|00\rangle+\sqrt{0.25}|11\rangle+\sqrt{0.25}|22\rangle.$

  It is obvious that $|\psi_1\rangle$ and $|\psi_2\rangle$ are
  incomparable even under ELOCC \cite{JP99}. Furthermore, the maximal conversion probability \cite{Vidal99}
  is given by
       $$P_{max}(|\psi_1\rangle^{\otimes k}\rightarrow |\psi_2\rangle^{\otimes k})=0.8^k,$$
which is  exponentially  decreasing  when $k$ increases
\cite{SRS02}.

  On the other hand, we have
  $R(|\psi_1\rangle,|\psi_2\rangle)=1$. By Theorem \ref{measure},
  it  holds  that $D(|\psi_1\rangle,|\psi_2\rangle)\leq 1$.  A
  numerical calculation leads to $$|\psi_1\rangle^{\otimes m}\rightarrow
|\psi_2\rangle^{\otimes m-1}$$
  for each $m=2,3,\ldots, 100$, which yields
  $$D(|\psi_1\rangle,|\psi_2\rangle)\geq 0.99.$$
That is, by a probabilistic manner, we  have only a very small
probability to obtain $100$ copies of $|\psi_1\rangle$ from the
same number of $|\psi_2\rangle$; while by a deterministic manner,
we can  obtain  $99$ copies of $|\psi_2\rangle$ from $100$ copies
of $|\psi_1\rangle$.\hfill $\square$

\end{example}
\section{conclusion}

To summarize, we introduce the notion of deterministic
entanglement exchange rate to characterize the degree of
convertibility of two entangled pure states. This quantity has a
very clear intuitive meaning: it denotes the optimal ratio of the
number of copies of target state and source state  under
deterministic LOCC . We prove that this rate is always positive,
and it is bounded from top by entropy ratio. In the special case
that the target state is maximally entangled, this upper bound can
be achieved. We  further prove that even allowing extra states to
serve as catalysts, the deterministic entanglement exchange rate
cannot be increased. We give some concrete examples to illustrate
the application of the main results. Especially, we demonstrate
that there can be two incomparable entangled states with
deterministic entanglement exchange rate larger than one.  We also
show that a catalyst can help us in obtaining more precise lower
bounds of deterministic entanglement exchange rate.

There are still many open problems for further studies. The most
interesting one is to determine the achievability of the upper
bound of deterministic entanglement exchange rate. We believe that
such an upper bound can always be achieved in general.

\textbf{Acknowledgement:} The authors wish to thank X. Li for
helpful discussions about majorization. The acknowledgement is
also given to the colleagues in the Quantum Computation and
Quantum Information Research Group. This work was partly supported
by the Natural Science Foundation of China (Grant Nos: 60273003,
60433050, 60321002, and 60305005). Especially, Runyao Duan
acknowledges the financial support of a PhD Student Creative
Foundation of Tsinghua University (Grant No: 052420003).
\smallskip\

\end{document}